# Adhesive contact between a rigid body of arbitrary shape and a thin elastic coating

Valentin L. Popov

Technische Universität Berlin, Str. des 17. Juni 135, 10623 Berlin, Germany

**Abstract.** Application of the principle of energy balance to a rigid indenter in contact with elastic layer on a flat rigid substrate provides a very simple derivation of the detachment criterion which earlier has been obtained by much more complicated asymptotic analysis. This criterion allows calculating the adhesive strength of arbitrary contact of a flat-ended indenter, which occurs to be proportional to the area of the face of the indenter and does not depend on its shape. Similarly, the adhesive contact problem can be easily solved in the case of arbitrary three-dimensional shape.

## 1. Introduction

A. Papangelo considered in [1] adhesion of an axially-symmetric rigid indenter with a thin elastic layer. In the present paper, we show how this problem can be solved for an indenter of arbitrary, not necessarily axis-symmetric, shape.

The problem which we consider is the following: A rigid body is indented into a thin elastic layer glued to the rigid flat substrate. We assume that the contact size (in the case of axially symmetrical profiles contact radius) is much larger than the thickness $h_0$ of the layer. In this case, one can assume that the elastic layer is deformed uni-axially, independently in each point. It can be considered as a two-dimensional elastic foundation with effective modulus [2]

$$\tilde{E} = \frac{E(1-\nu)}{(1+\nu)(1-2\nu)}, \tag{1}$$

where $E$ is elastic modulus and $\nu$ Poisson number. The layer can be represented as a two-dimensional elastic foundation consisting of independent springs which are placed with separation $\Delta x$ and $\Delta y$ in the directions $x$ and $y$ correspondingly, while each spring has the stiffness

$$\Delta k = \tilde{E} \frac{A}{h_0} \tag{2}$$

with $A = \Delta x \Delta y$.

Now consider a contact with a rigid profile $f(x,y)$ indented by $d$. The displacements of the springs in contact is given by

$$u_z(x,y) = d - f(x,y). \tag{3}$$

## 2. Detachment criterion

The boundary of the contact is determined by the criterion of energy balance according to Griffith [3], similarly as how this has been done in [4]. If the elongation of a spring at the boundary of the contact area, due to adhesion, is $\delta l$, then detachment will lead to relaxation of elastic energy

$$\Delta U_{el} = \frac{1}{2}\Delta k \cdot \delta l^2 = \frac{1}{2}\tilde{E}\frac{A}{h_0}\delta l^2. \tag{4}$$

The critical elongation is given by the condition that this elastic energy is equal to the work of adhesion of this cell,

$$U_{ad} = \gamma_{12} A \tag{5}$$

where $\gamma_{12}$ is specific work of separation. The condition for detachment has the form

$$\frac{1}{2}\tilde{E}\frac{A}{h_0}\delta l_c^2 = \gamma_{12}A, \tag{6}$$

which provides for critical elongation

$$\delta l_c = \sqrt{2\gamma_{12}\frac{h_0}{\tilde{E}}}. \tag{7}$$

Thus, using (3), the equation determining the boundary is given by

$$d - f(x,y) = -\delta l_c = -\sqrt{2\gamma_{12}\frac{h_0}{\tilde{E}}} \tag{8}$$

or

$$f(x,y) = d + \sqrt{2\gamma_{12}\frac{h_0}{\tilde{E}}} \tag{9}$$

which corresponds to a cross-section of the profile $f(x,y)$ at the height $d + \sqrt{2\gamma_{12}h_0/\tilde{E}}$. This result was found earlier by asymptotic analysis [5] but follows also directly from the application of the Griffith energy balance.

## 3. Detachment of flat-ended punches

Consider a flat-ended punch of arbitrary shape. If the punch is bought into contact with the layer and then lifted, the detachment criterion (7) will be fulfilled at the same time for all elements at the boundary of the punch, so that the force in this moment will be

$$F_N = -\tilde{E}\frac{A_{\text{punch}}}{h_0}\sqrt{2\gamma_{12}\frac{h_0}{\tilde{E}}} = -A_{\text{punch}}\sqrt{\frac{2\tilde{E}\gamma_{12}}{h_0}}, \tag{10}$$

where $A_{\text{punch}}$ is the total area of the punch. The adhesive strength depends only on the area of the punch but not on its particular shape.

## 4. Detachment of a power-law profile

Let us consider an axially-symmetric indenter having the form

$$f(r) = \frac{r^k}{kR^{k-1}} \tag{11}$$

(considered also in the paper [1]). The contact radius $a$ is given by the condition

$$\frac{a^k}{kR^{k-1}} = d + \sqrt{2\gamma_{12}\frac{h_0}{\tilde{E}}} \tag{12}$$

or

$$a = \left[kR^{k-1}\left(d + \sqrt{2\gamma_{12}\frac{h_0}{\tilde{E}}}\right)\right]^{1/k} \tag{13}$$

The normal force is given by

$$F_N = \frac{\tilde{E}}{h_0}\int_0^a 2\pi r\left(d - \frac{r^k}{kR^{k-1}}\right)dr = 2\pi\frac{\tilde{E}}{h_0}\left[d\cdot\frac{a^2}{2} - \frac{a^{k+2}}{k(k+2)R^{k-1}}\right]$$

$$= 2\pi\frac{\tilde{E}}{h_0}\left[\left(\frac{a^k}{kR^{k-1}} - \sqrt{2\gamma_{12}\frac{h_0}{\tilde{E}}}\right)\frac{a^2}{2} - \frac{a^{k+2}}{k(k+2)R^{k-1}}\right] \tag{14}$$

$$= \pi\frac{\tilde{E}}{h_0}\left(\frac{a^{k+2}}{(k+2)R^{k-1}} - a^2\sqrt{2\gamma_{12}\frac{h_0}{\tilde{E}}}\right)$$

This force achieves extremum at

$$a_c = \left(2R^{k-1}\sqrt{2\gamma_{12}\frac{h_0}{\tilde{E}}}\right)^{1/k} \tag{15}$$

The force of adhesion is given by

$$F_A = 2^{\frac{k+6}{2k}}\pi\frac{k}{(k+2)}\left(\frac{\tilde{E}}{h_0}\right)^{\frac{k-2}{2k}} R^{\frac{2(k-1)}{k}} \gamma_{12}^{\frac{k+2}{2k}}. \tag{16}$$

In particular, for a parabolic profile, $F_A = 2\pi R\gamma_{12}$ which coincides with the result obtained in [5] and [1].

## 5. Conclusion

Application of the Griffith principle of energy balance to an adhesive contact of a rigid indenter and an elastic layer leads to a simple detachment criterion stating that the critical elongation in all points of the contact boundary is constant and equal to $\delta l_c = \sqrt{2\gamma_{12}\frac{h_0}{\tilde{E}}}$. This means that the contact profiles is simply provided by the cross-section of indenter at the height $d + \sqrt{2\gamma_{12}h_0/\tilde{E}}$. This solves the adhesive problem in the general case of arbitrary indenter shape.